\documentclass[prc,aps,epsfig]{revtex4}
\usepackage{graphicx}
\usepackage{epsfig}
\def\beq{\begin{equation}}
\def\enq{\end{equation}}
\def\beqa{\begin{eqnarray}}
\def\enqa{\end{eqnarray}}
\def\lb{\label}
\def\nn{\nonumber}
\def\MeV{\nobreak\,\mbox{MeV}}
\def\GeV{\nobreak\,\mbox{GeV}}
\def\pli{p^\prime}
\newcommand{\rag}{\rangle}
\newcommand{\lag}{\langle}

\begin{document}

\title{Looking for meson molecules in B decays}

\author{I. Bediaga}

\affiliation{Centro Brasileiro de Pesquisas F\'\i sicas, Rua Xavier 
Sigaud 150, 22290-180 Rio de Janeiro, RJ, Brazil}

\author{F.S. Navarra, M. Nielsen}
\affiliation{Instituto de F\'{\i}sica, Universidade de S\~{a}o Paulo\\
 C.P. 66318,  05315-970 S\~{a}o Paulo, SP, Brazil}

\begin{abstract}

We discuss the possibility of observing a loosely bound molecular state in a 
B three-body hadronic decay. In particular we use the QCD sum rule approach to study
a $\eta^\prime-\pi$ molecular current. We consider an isovector-scalar $I^G~J^{PC}=
1^-~0^{++}$ molecular current and we use the two-point and three-point functions 
to study the mass and decay width of such state. We consider the contributions of 
condensates up to dimension six and we work at leading order in $\alpha_s$. 
We obtain a mass  around 1.1 GeV, consistent with a loosely bound state, and a 
$\eta^\prime-\pi\rightarrow K^+ K^-$ decay width around 10 MeV.

\end{abstract}



\maketitle



\vspace{1cm}
\section{Introduction}

One of the outstanding open questions in hadron physics is: are there 
meson-meson bound states? The same mechanism of meson exchange that binds the 
deuteron could also in principle bind two mesons. The interest in this subject 
was renewed by the discovery of the  new charmonium states.  
Since their first appearance, some of them were considered to be meson 
molecules. These states have already been discussed in some reviews 
\cite{Brambilla:2010cs,Nielsen:2009uh,Olsen:2009gi}. 

In this note we discuss how to look for molecules at the LHCb taking  advantage 
of the unprecedented high statistics.

We can look for  meson molecules in three-body hadronic $B$ decays. Since the
 phase space is large we can even try 
to use directly the Dalitz diagram, which  extends up to large values of the 
variables $s_{12}$ and $s_{23}$.
All the known normal quark-antiquark  intermediate resonant states, leave an 
imprint in  the Dalitz plot, which is directly related with the quantum numbers
of the states and lead to the identification of the state. Examples are:
a continuous straight line, in the case of scalar states, a line with a 
hole, in the case of vector states, or a line with two holes, in the case
of tensor states. 

\begin{figure}[h]
\begin{center}
\epsfig{file=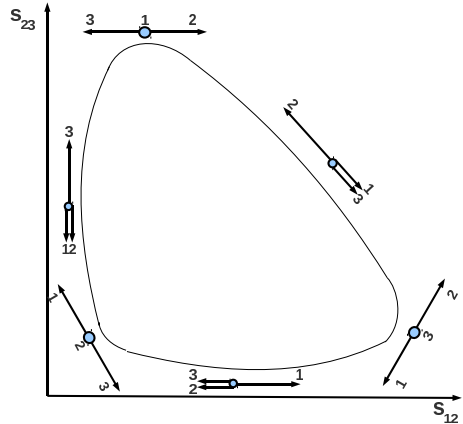,width=6cm}
\caption{Dalitz plot of a three-body $B$ decay. The small drawings illustrate 
the different kinematical configurations.}
\end{center}
\label{fig1}
\end{figure}

A sketch of a Dalitz plot for a three-body  meson decay is shown in 
Fig. \ref{fig1}, where for each invariant parameter $s_{12}$ or $s_{23}$
it is shown the relative momentum of each one of the two particles $1,~2$ or $2,~3$.
Let us consider the case that the particles $1$ and $2$ are pions coming from the 
$\rho$ meson decay. Of course this decay should produce a line parallel to the 
$s_{23}$ axis in the point $s_{12}=m_\rho^2$. However, since the pions coming from the
$\rho$ meson decay must have one unit of angular momentum, they cannot go both
to the same direction. Therefore, no pions could be seen in the region where
the relative momentum between them is small. From  Fig.\ref{fig1} one  can see 
that this region is just in the middle of the line parallel to the $s_{23}$ axis. 
Therefore, a line characterizing a 
vector resonance state must have a hole in the middle, as mentioned above.
Now imagine that the resonant state is a loosely bound molecular state of the 
particles $1,~2$. 
A loosely bound molecular state can only exist when the relative momentum between
the two mesons in the molecule is small. In
this case one has exactly the opposite situation than the one discussed before:
there will be no signal in the Dalitz plot unless the two mesons in the
molecular state  go in the same diretion. Or, in other words, one expect
a small line parallel to the $s_{23}$ axis in the middle of the Dalitz plot, 
approximately in the region where there is a hole in the line characterizing a 
vector resonance. 

\begin{figure}[t]
\begin{tabular}{cc}
\includegraphics[scale=0.22]{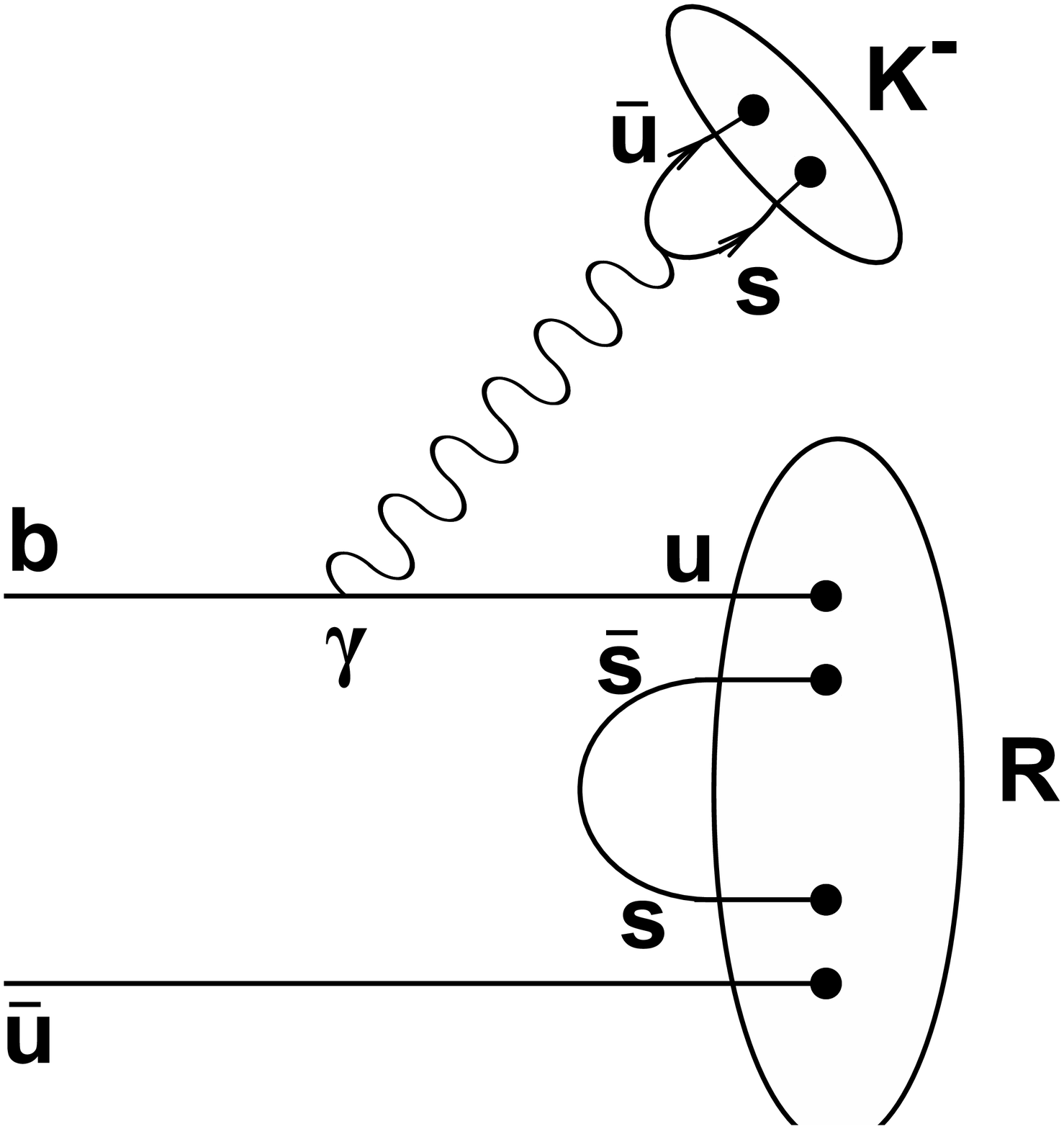}
\hspace{0.5cm}
\includegraphics[scale=0.22]{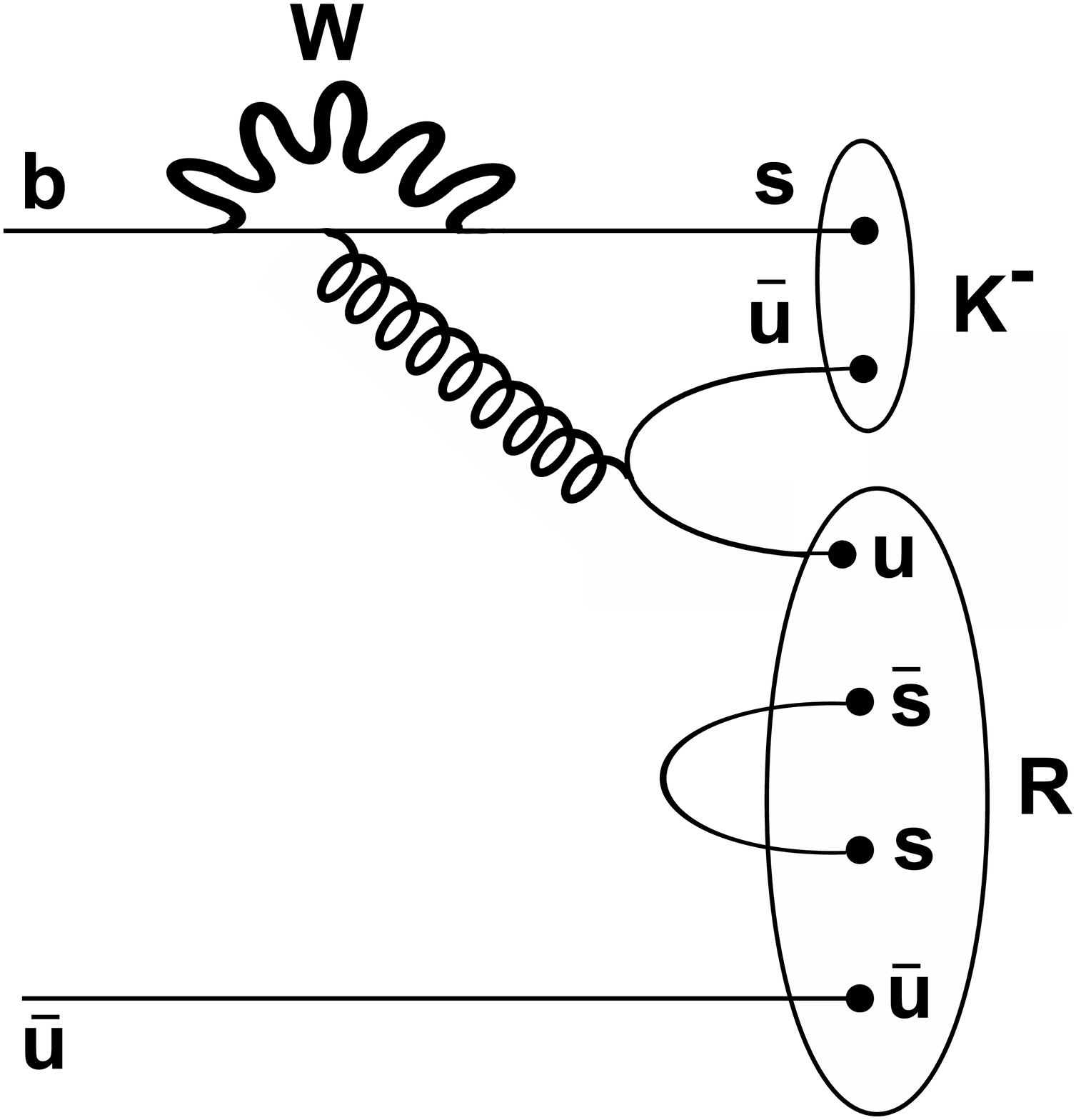}
\end{tabular}
\caption{ The two  relevant diagrams  for the $B^- \rightarrow K^+K^- K^- $, 
through the resonance $R$. }
\label{fig2}
\end{figure}

\begin{figure}[t]
\begin{tabular}{cc}
\includegraphics[scale=0.22]{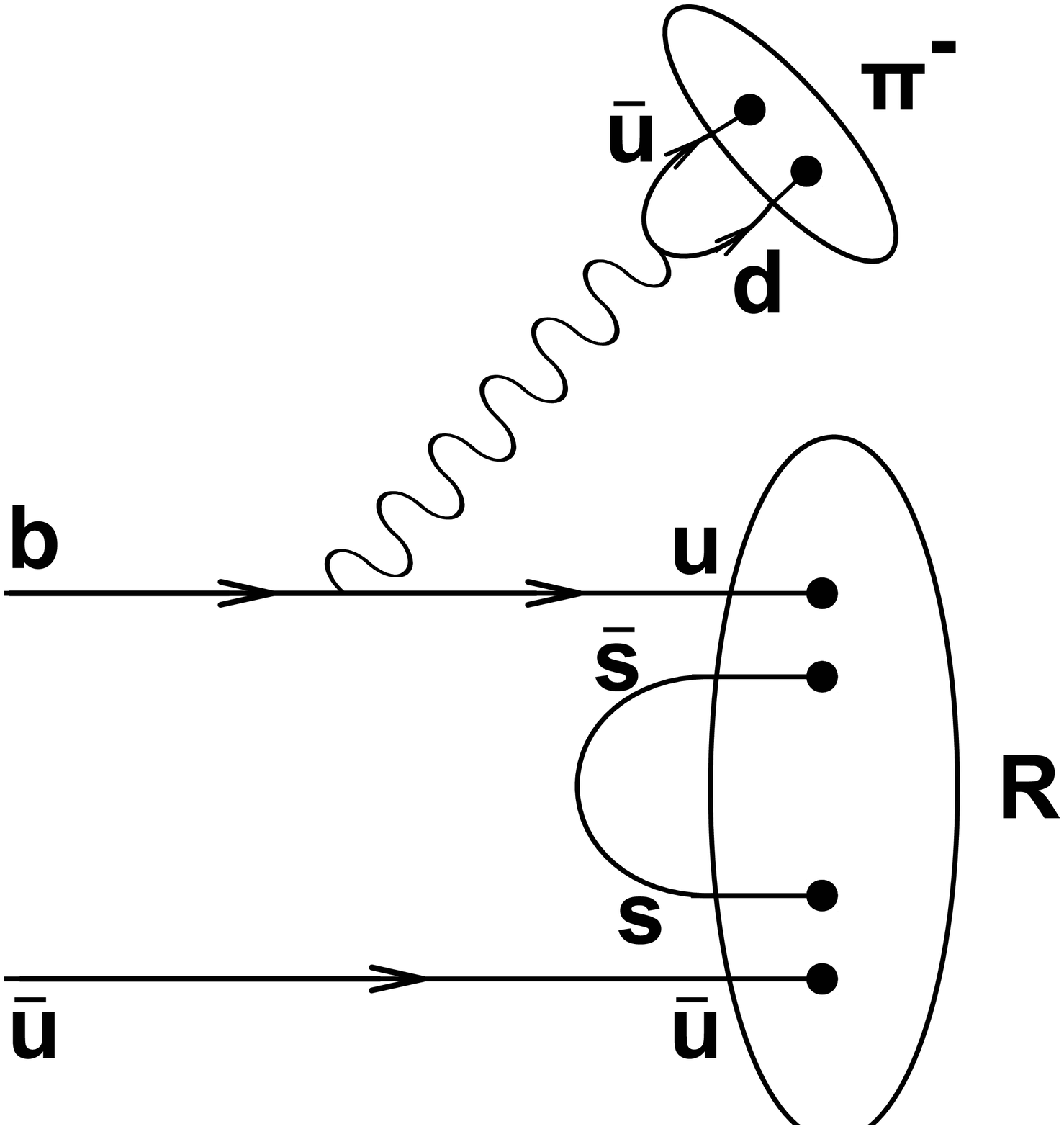}
\vspace{1cm}
\includegraphics[scale=0.22]{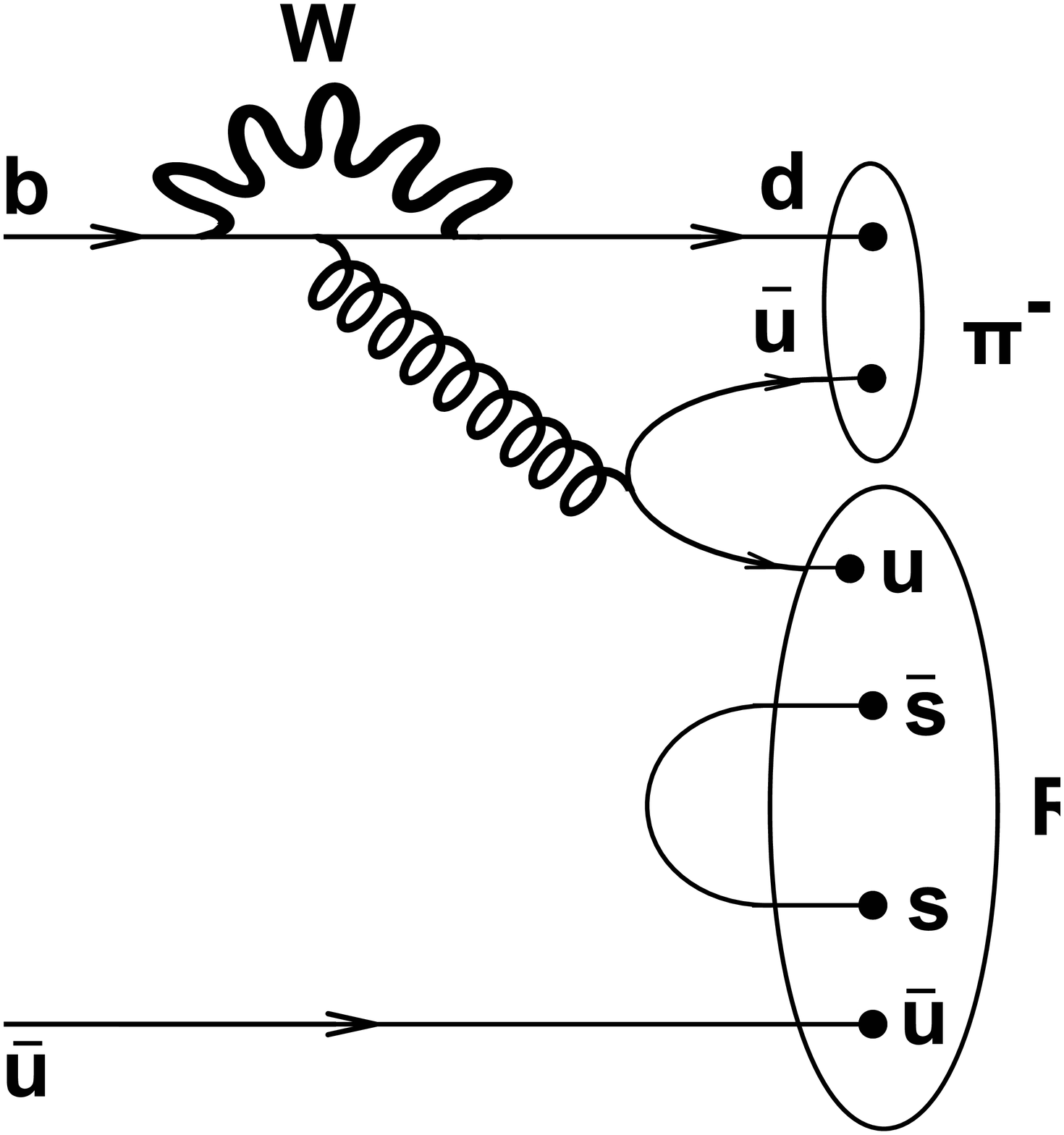}
\end{tabular}
\caption{ The two  relevant diagrams  for the $B^- \rightarrow K^+K^- \pi^- $ decay,
through the resonance $R$. }
\label{fig3}
\end{figure}

The final particles observed in the three-body $B$ decays are pions and kaons. 
Therefore, to observe a molecular state in the Dalitz plot for a three-body $B$ 
decay, this molecular state must decay into pions and/or kaons. Let us consider 
a $\eta' - \pi$  loosely bound molecule with the quark content 
$\bar{u} u \bar{s} s$. This resonant state, hereafter called $R$,  is especially 
interesting because its mass,  $m_R$, should be approximately given by:
\begin{equation}
m_R \sim m_{\eta'} + m_{\pi} = 958 + 138 = 1096  \, \mbox{GeV}
\label{massmoc}
\end{equation}
and, therefore, it is quite visible in the $B$ decay Dalitz plot. Since for $S$-wave
this molecule has $I^GJ^{PC}=1^-0^{++}$, it can not decay into $\pi^+ \pi^-$, but
it will decay into $K^+ K^-$. In particular, there 
are already data for $B^- \rightarrow K^+ K^-  K^- $, $B^- \rightarrow K^+ K^- 
 \pi^- $, $B^- \rightarrow \pi^+\pi^-\pi^- $  and $B^-\rightarrow\pi^+\pi^- K^- $
\cite{pdg}.
Only in the first two of these cases the decay could go through  the resonant 
state $R$, 
as illustrated in Figs.~\ref{fig2} and \ref{fig3}. For these cases, 
a small line with $\sqrt{s_{12}}\sim 1.1\mbox{ GeV}$ parallel to the 
$s_{23}$ axis should be seen in the Dalitz plot, in the region where the two particles
$\eta'$ and $\pi$ have a small relative momentum. This signal should be very 
different from all other established resonant states decaying into  $K^+K^-$,
like the $a_0$ for instance, and should be only seen in the channels
$B^- \rightarrow K^+ K^-  K^- $ and $B^- \rightarrow K^+ K^- \pi^- $.
Of course, the figure is very qualitative and it is not 
possible to say how large is the line segment around the indicated position. 
However, the observation of this structure in the Dalitz plot of the two 
mentioned $B$ decays, and not in the others, would represent a strong evidence 
of the formation of this molecular state.  The observation of a line that only 
appears in a certain piece of the $s_{23}$ axis with a fixed value of $s_{12}$, and
only for the decays $B^- \rightarrow K^+ K^-  K^- $ and $B^- \rightarrow K^+ K^- 
\pi^- $, could be interpreted as the existence of  a weakly bound molecular state.

\section{The $\eta' - \pi$  scalar molecule}

\subsection{Mass} 

In a previous work \cite{nosca}  we have investigated the possiblity that the light 
scalar states could be interpreted  as tetraquark states.  Now we perform a 
complementary investigation, trying to understand a possible $\eta'\pi$ meson 
molecular state  in the QCD sum rule (QCDSR) framework \cite{svz,rry,SNB}.

The QCDSR approach is based on the correlator of hadronic currents. 
A generic two-point correlation function is given by

\beq
\Pi(q)=i\int d^4x ~e^{iq.x}\lag 0| T [j(x)j^\dagger(0)] |0\rag,
\lb{2po}
\enq
where the local current $j(x)$ contains all the information about the hadron of
interest, like quantum numbers, quarks content and so on. A molecular current can be
constructed  from the mesonic currents that describe the two mesons in the molecule.
In the case of scalar $\eta' - \pi$ state a possible current is:
\beq
  j= \left( \bar{u}_i \:\gamma_5 \:u_i-\bar{d}_i \:\gamma_5 \:d_i \over\sqrt{2}
\right)\left[\sin\theta\left( \bar{u}_j \:\gamma_5 \:u_j+\bar{d}_j \:\gamma_5 \:d_j 
\over\sqrt{2}\right)+\cos\theta~(\bar{s}_j\gamma_5\:s_j)\right],
  \label{pse}
\enq
where $i,j$ are color indices, $u,~d,~s$ are the up down and strange quark 
fields respectively, and the mixing angle, $\theta$, in the $\eta^\prime$ current
is $\theta\sim40^0$ \cite{Anisovich:1995sq,Feldmann:1998vh,Balakireva:2012um}.
In this work we use $\theta=40^0$.

In general,  there is no one to one correspondence between
the current and the state, since a molecular current
can be rewritten in terms of a sum over tetraquark type
currents through a Fierz transformation.  However, as shown in 
\cite{Nielsen:2009uh}, if the physical state is a molecular state, it would be 
better to choose a molecular type of current so that it has a large overlap 
with the physical state. In any case, it is very important to notice that since
the current in Eq.~(\ref{2po}) is local, it does
not represent an extended object, with two mesons separated in space, but
rather a very compact object with two singlet quark-antiquark pairs. 

The coupling of the scalar resonance $R$, to the scalar current $j$,  can be
parametrized in terms of a parameter $\lambda$ as:
\beq
\lag 0 | j|R\rag =\lambda\;.
\lb{lam}
\enq

In the QCD side evaluation of the correlator function in Eq.(\ref{2po}) we work at 
leading order and consider condensates up to 
dimension six. We deal with the strange quark as a light one and consider
the diagrams up to order $m_s$. We neglect the terms proportional to $m_u$ and $m_d$.
In the phenomenological side we consider
the usual pole plus continuum contribution. Therefore, we introduce
the continuum threshold parameter $s_0$ \cite{io1}. In the $SU(2)$ limit the
quarks $u$ and $d$ are degenerate and we consider the $u$-quark condensate equal
to the $d$-quark condensate, wich we call $\langle\bar{q}q\rangle$. After doing a 
Borel tranform in both sides of the calculation the sum rule is given by:
\beqa
\lambda^2e^{-m_R^2/M^2}&=&3{M^{10}E_4\over 2^{13}5\pi^6}(12+\sin^2{\theta})-
{m_s\langle\bar{s}s\rag M^6E_2\over 2^7\pi^4}\cos^2{\theta}+
{\lag g^2G^2\rag M^6E_2\over2^{13}\pi^6}(4-\sin^2{\theta})
\nn\\
&-&{m_s\lag\bar{s}g\sigma.Gs\rag\over 2^7\pi^4}
M^4E_1\cos^2{\theta}\left(3.5-3\ln(M^2/
\Lambda_{QCD}^2)\right)\nn\\
&+&{M^4E_1\over2^6\pi^2}\left(\lag\bar{q}q\rag^2(1+3\cos^2{
\theta})+2\lag\bar{s}s\rag^2\sin^2{\theta}\right),
\label{sr}
\enqa
where
\beq
E_n\equiv 1-e^{-s_0/M^2}\sum_{k=0}^n\left(s_0\over M^2\right)^k{1\over k!}
\ ,
\label{con}
\enq
which accounts for the continuum contribution. 

In the numerical analysis of the sum rules, the values used for the
quark masses and condensates are \cite{narpdg,ri07}: 
$m_s = 0.13\,\GeV$,
$\lag\bar{q}q\rag=\,-(0.23)^3\,\GeV^3$, $\lag\bar{s}s\rag=\,0.8\lag\bar{q}q\rag$,
$\lag\bar{q}g\sigma.Gq\rag=m_0^2\lag\bar{q}q\rag$ with \cite{SNB}
$m_0^2=0.8\,\GeV^2$ and 
$\lag g^2G^2\rag=0.88~\GeV^4$.

\begin{figure}[h] 
\centerline{\epsfig{figure=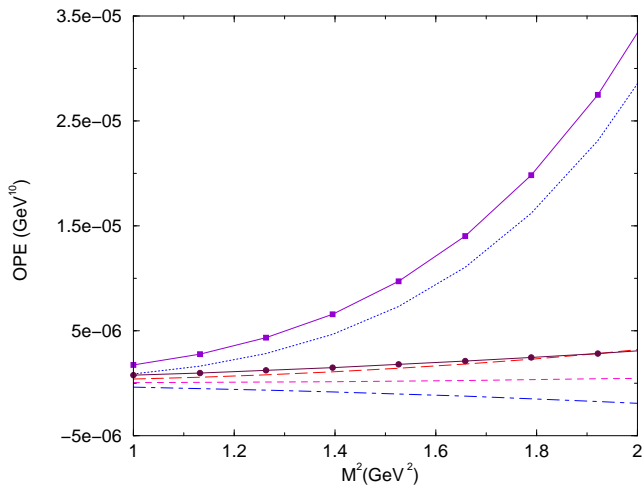,height=65mm}}
\caption{The OPE convergence in the region $1.0 \leq M^2 \leq
2.0~\GeV^2$ for $\sqrt{s_0} = 1.5 \GeV$. The dotted, dashed, long-dashed, dot-dashed,
solid with circles and solid whit squares lines give the perturbative, quark
condensate, gluon condensate, mixed condensate, four-quark condensate and total 
contributions repectively.}
\label{fig4} 
\end{figure}

In Fig.~\ref{fig4} we show the OPE convergence  of the sum rule
in Eq.~(\ref{sr}). From this figure we see that the  convergence is  reasonable 
for $M^2>1.2~\GeV^2$ and  very good  for $M^2>1.5~\GeV^2$. 
However, as in the case of the light scalars \cite{ri07},
there is no pole dominance for these values of $M^2$. This result could be 
interpreted in two different ways: i) it could indicate that this state does not 
exist, or ii) it could indicate that this state is not clearly separated from
the continuum. The second interpretation can be applied to  very broad states,
as the light scalars $\sigma$ and $\kappa$, since their widths  are as large as
the difference between their masses and the continuum threshold. In what follows we  
stick to the  interpretation ii).

\begin{figure}[h] 
\centerline{\epsfig{figure=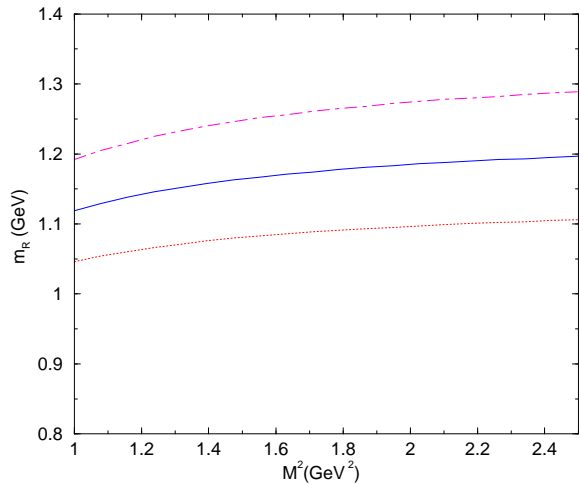,height=65mm}}
\caption{The resonance mass as a function of the sum rule parameter
($M^2$) for different values of the continuum threshold: $\sqrt{s_0} = 1.4 \GeV$ 
(dotted line), $\sqrt{s_0} = 1.5 \GeV$
(solid line) and $\sqrt{s_0} = 1.6 \GeV$ (dot-dashed line).}
\label{fig5} 
\end{figure} 
In order to extract the mass $m_{R}$ without knowing   the value of
the constant $\lambda$, we take the derivative of Eq.~(\ref{sr})
with respect to $1/M^2$ and divide the result by Eq.~(\ref{sr}). 
In Fig.~\ref{fig5}, we show the resonance mass as a function of $M^2$ for different 
values of $\sqrt{s_0}$. We limit ourselves to the region $M^2>1.2~\GeV^2$
where the curves are more stable and where the OPE convergence is better. 
Averaging the mass over all this region we find:
\beq
m_R = (1.15\pm 0.10) \GeV~,
\label{msr}
\enq
which is compatible with the experimental threshold in Eq.(\ref{massmoc}).
Having the mass, we can also evaluate the value of the parameter $\lambda$ that
gives the coupling between the state and the current. We obtain:
\beq
\lambda = (1.39\pm 0.27)\times10^{-3} \GeV^5~.
\enq

\subsection{Decay width}

In order to study the $RK^+K^-$ vertex associated with
the $R\rightarrow K^+K^-$ decay, we consider the three-point function
\beq
T_{\mu\nu}(p,\pli,q)=\int d^4x~d^4y~e^{i.\pli.x}~e^{iq.y}\lag0|T\{
j_{5\mu}^{K^+}(x)j_{5\nu}^{K^-}(y)j^\dagger(0)\}|0\rag,
\lb{3point}
\enq
where $p=\pli+q$, $j$ is given in Eq.~(\ref{pse}) and we use the axial currents
for the kaons:
\beq
j_{5\mu}^{K^+}=
\bar{s}_a\gamma_\mu\gamma_5u_a,\,\;\;\;j_{5\mu}^{K^-}=
\bar{u}_a\gamma_\mu\gamma_5s_a.
\lb{pseu}
\enq
To evaluate the phenomenological side
we insert intermediate states for $K^+$, $K^-$ and $R$, and we use the 
definitions 
in Eqs.~(\ref{lam}) and (\ref{fp}) bellow:
\beq
\lag 0 | j_{5\mu}^{K}|K(p)\rag =ip_\mu F_{K}\;.
\lb{fp}
\enq
We obtain the following relation: 
\beq
T_{\mu\nu}^{phen} (p,\pli,q)={F_{K}^2\lambda
\over (M_{R}^2-p^2)(m_{K}^2-{\pli}^2)(m_{K}^2-q^2)}~g_{RKK}~\pli_\mu 
q_\nu
+\mbox{  higher resonances}\;,
\lb{phen}
\enq
where the coupling constant $g_{RKK}$ is defined by the matrix element:
\beq
\lag K(\pli)K(q)|R(p)\rag=g_{RKK}.
\enq

Here we follow refs.~\cite{nosca,nari} and work at the kaon pole, as suggested
in \cite{rry} for the nucleon-pion coupling constant.
This method was also applied to the nucleon-kaon-hyperon coupling 
\cite{ccl,brann}, to the $D^*-D-\pi$ coupling \cite{nnbcs,ppnp} and to the
$J/\psi-\pi$ cross section \cite{nnmk}. It consists in neglecting the kaon mass
in the denominator of Eq.~(\ref{phen}) in the term ${1/(m_{K}^2-q^2)}$, and 
working at $q^2=0$. In the QCD side
one singles out the leading terms in the operator product expansion of 
Eq.(\ref{3point}) that match the $1/q^2$ term. Up to dimension six only
the diagrams proportional to the quark condensate times $m_s$ and the 
four-quark condensate contribute.
Making a single Borel transform to both $-p^2=-{\pli}^2\rightarrow M^2$ we get:
\beqa
g_{R K^+K^-}{\lambda F_K^2\over m_{R}^2-m_K^2}
(e^{-m_K^2/M^2}-e^{-m_R^2/M^2})&=&{\sqrt{2}\cos{\theta}\over8}\left({(\lag\bar{q}q\rag
+\lag\bar{s}s\rag)^2\over3}+\right.
\nn\\
&+&\left.{m_s\over8\pi^2}
\left(\lag\bar{q}q\rag-{\lag\bar{s}s\rag\over3}\right) M^2
\left(1-e^{-s_0^K/M^2}\right)\right),
\lb{sr3}
\enqa
where $s_0^K=(1.0\pm0.1)\GeV^2$, is the continuum threshold for the kaon.

As discussed in ref.~\cite{io2}, the problem of doing a single Borel transformation, 
in a three-point function sum rule, is the fact that terms
associated with the pole-continuum transitions are not suppressed. However, as shown in
\cite{io2}, the pole-continuum transition term 
has a different behavior, as a function of the Borel mass, as compared with
the double pole contribution: it grows with $M^2$. Therefore, the 
pole-continuum contribution can be taken into account through the introduction
of a parameter $A$ in the phenomenological side of the sum rule in 
Eq.~(\ref{sr3}), by making the substitution
$g_{RK^+K^-}\rightarrow g_{RK^+K^-}+AM^2$ \cite{nosca,brann,nnbcs,nnmk}.

\begin{figure}[h] 
\centerline{\epsfig{figure=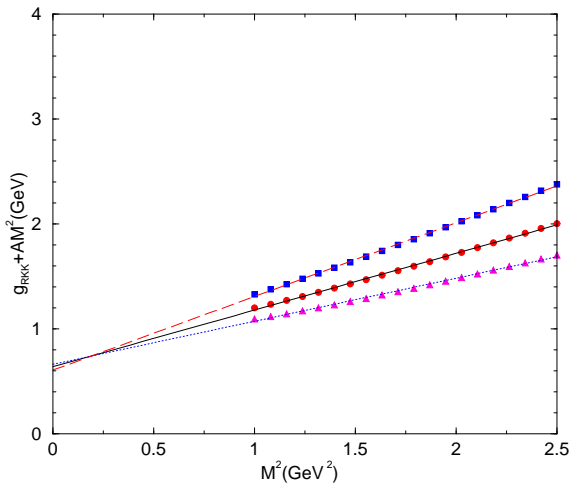,height=65mm}}
\caption{The QCDSR result for the coupling constant $g_{RK^+K^-}$ as a 
function of the sum rule parameter $M^2$ for different values of ${s_0}$ 
and $s_0^K$ (circles, triangles and squares for 
$\sqrt{s_0}=1.5\GeV,~s_0^K=1.0\GeV^2;~\sqrt{s_0}=1.6\GeV,~s_0^K=1.1\GeV^2$
and $\sqrt{s_0}=1.4\GeV,~s_0^K=0.9\GeV^2$ respectively).
The solid, dotted and dashed lines give the linear fit to the 
QCDSR results. }
\label{fig6} 
\end{figure} 

Using $F_K=160\MeV$, $m_K=490\MeV$, $m_R=1.15\GeV$ and the parameter 
$\lambda$ given by the sum rule in Eq.~(\ref{sr})
we show, in Fig.~\ref{fig6}, the QCDSR results for the vertex coupling constant,
for different values of $s_0$ and $s_0^K$ in the interval given above.
We see that, in the Borel range used for the two-point function, the QCDSR
results do have  a linear behaviour as a function of the Borel mass. Fitting
the QCDSR results by a linear form: $g_{RK^+K^-}+AM^2$ (which is also shown
in Fig.~\ref{fig6}), the coupling can be obtained by extrapolating the fit 
to $M^2=0$. In the  limits of the continuum thresholds mentioned above 
and taking into account the uncertainties in $m_R$ given in Eq.~(\ref{msr}) we obtain: 
\beq
g_{R K^+K^-}=(0.63 \pm 0.06)\GeV.
\lb{coup}
\enq

The decay width of $R\rightarrow K^+K^-$ is given in terms of the hadronic
coupling $g_{RK^+K^-}$ as: 
\beq
\Gamma(R\rightarrow K^+K^-)={1\over 16\pi m_R^3}g_{RK^+K^-}^2\sqrt{\lambda(
m_R^2,m_{K}^2,m_{K}^2)},
\lb{decay}
\enq
where $\lambda(m_R^2,m_{K}^2,m_{K}^2)=m_R^4+m_{K}^4+m_{K}^4-2m_R^2
m_{K}^2-2m_R^2m_{K}^2-2m_{K}^2m_{K}^2=m_R^2(m_R^2-4m_K^2)$.
Therefore, we get:
\beq
\Gamma(R\rightarrow K^+K^-)=(11.4\pm2.2)\MeV.
\enq
Of course this is not the total width of the $\eta^\prime\pi$ molecule, since
it can also decay into $\eta-\pi$ with a much bigger phase space. However, in the
$B$ decays discussed here, only the channel $R\rightarrow K^+K^-$ can be observed.

The errors quoted above  come directly from the uncertainty in the determination
of the continuum threshold parameters, $s_0$. According to our previous experience, 
they are the main source of uncertainty in the method. For a detailed analysis of 
the uncertainty associated to other parameters used in QCDSR we refer the reader to 
Refs. \cite{Nielsen:2009uh} and \cite{ppnp}.

\section{Conclusion}

We have proposed that a loosely bound molecular state should leave a particular 
signal in the Dalitz plot. A loosely bound molecular state, of the particles $1,~2$,  
can only exist when the relative momentum between these two particles  
is small. Therefore, we expect to observe  a short line parallel to the $s_{23}$ axis in the 
middle of the Dalitz plot, approximately in the region where there is a hole in the 
line characterizing a vector resonance (see Fig.~\ref{fig1}). This signal is 
different from any signal charactering the normal quark-antiquark mesons, and could be used 
to identify the existence of loosely bound molecular states. 

In the case of three-body $B$ decays, the final particles observed are pions and 
kaons. Therefore, to observe a molecular state in the Dalitz plot for a 
three-body $B$ decay, this molecular state must decay into pions and/or kaons. 
We have considered a $\eta' - \pi$ molecular state. If this state exists as
a loosely bound state, its mass should be close to the $\eta' - \pi$ threshold:
$\sim1.1\GeV$, quite visible in the $B$ decay Dalitz plot.
Since for a $S$-wave this molecule has $I^GJ^{PC}=1^-0^{++}$, it can 
not decay into $\pi^+ \pi^-$, but it will decay into $K^+ K^-$. Therefore,
the observation of a  small line with $\sqrt{s_{12}}\sim 1.1\mbox{ GeV}$ parallel 
to the $s_{23}$ axis in the Dalitz plot for the $B^- \rightarrow K^+ 
K^-  K^- $ and $B^- \rightarrow K^+ K^- \pi^- $ decays, with negative observation in 
the Dalitz plot for the $B^- \rightarrow \pi^+ \pi^-  K^- $ and $B^- \rightarrow \pi^+
 \pi^- \pi^- $ decays would definitively indicate the existence of the
 $\eta' - \pi$ molecular state.

We have used QCD sum rules to study the mass and the decay width, of a
$\eta' - \pi$ molecular current, using two-point and three-point functions 
respectively.  We have considered diagrams up to dimension six in both cases. We 
found a mass a slightly larger than the $\eta' - \pi$ threshold, indicating the 
possiblity of a loosely bound molecular state. We obtained a small width for the 
$\eta' - \pi\rightarrow K^+ K^-$ decay around 10 MeV. With these informations 
should be possible to experimentaly indentify this state in the $B^- \rightarrow K^+ 
K^-  K^- $ and $B^- \rightarrow K^+ K^- \pi^- $ Dalitz plots, if it exists.

The method for the identification of resonances (or bound states) discussed here could be 
applied to other cases. A straightforward extension of our work could be done to the
 $\eta' - \pi$  with quantum numbers $ J^{PC} = 1^{-+}$. This exotic state has been recently 
searched for by  the CLEO \cite{cleo} and COMPASS \cite{compass} collaborations.

\begin{acknowledgments}
This work was  partially financed by the Brazilian funding agencies  CNPq and 
FAPESP.
\end{acknowledgments}

\end{document}